\documentclass[12pt]{article}
\usepackage{graphicx}
\newcommand{\beq}{\begin{equation}}
\newcommand{\eeq}{\end{equation}}
\newcommand{\bea}{\begin{eqnarray}}
\newcommand{\eea}{\end{eqnarray}}

\renewcommand{\a}{\alpha}

\begin{document}
\begin{titlepage}
\begin{flushleft}
       \hfill                      {\tt hep-th/0408040}\\
       \hfill                       FIT HE - 04-03 \\
\end{flushleft}
\vspace*{3mm}
\begin{center}
{\bf\LARGE Chiral symmetry breaking \\
driven by dilaton}

\vspace*{5mm}
\vspace*{12mm}
{\large Kazuo Ghoroku\footnote[2]{\tt gouroku@dontaku.fit.ac.jp}
and Masanobu Yahiro\footnote[3]{\tt yahiro2scp@mbox.nc.kyushu-u.ac.jp}
}\\
\vspace*{2mm}

\vspace*{2mm}

\vspace*{4mm}
{\large ${}^{\dagger}$Fukuoka Institute of Technology, Wajiro, 
Higashi-ku}\\
{\large Fukuoka 811-0295, Japan\\}
\vspace*{4mm}
{\large ${}^{\ddagger}$Department of Physics, Kyushu University, Hakozaki,
Higashi-ku}\\
{\large Fukuoka 812-8581, Japan\\}

\vspace*{10mm}
\end{center}

\begin{abstract}

Dynamical properties of gauge theories with light flavor quarks
are studied in a dual supergravity by adding a D7-brane probe into the 
AdS background deformed by the dilaton. By estimating the vev of 
flavor quark-bilinear in both the supersymmetric and non-supersymmetric
gravity duals, we find spontaneous chiral symmetry breaking in the case of 
the non-supersymmetric background. We also study quark-antiquark potential 
for light quarks to see the quark confinement in the models considered 
here. 

\end{abstract}
\end{titlepage}

\section{Introduction}

It is an interesting problem to make clear the gauge/gravity
correspondence from superstring theory \cite{MGW}. 
In particular, it is hoped that the correspondence 
could be applicable to QCD by deforming the anti-de Sitter space-time 
(AdS) into the non-conformal case.
It is however
a difficult problem to describe QCD in the ultra-violet (UV) region
where we must reproduce the asymptotic freedom and we will then need full
string theory to see this property. Then semi-classical gravity would not
be sufficient in this region. 

Meanwhile near the horizon, in the infrared (IR) region, the semi-classical
gravity could provide the characteristic property of QCD vacuum, e.g. quark-
confinement and chiral symmetry breaking etc.. Up to now, we could believe
that the most plausible vacuum state of QCD is in a gauge field 
condensate phase \cite{Shifman,Reinders}. 
When the vacuum expectation value (vev)
of gauge field exists, it could be seen through 
the dilaton configuration on the gravity side. So it is meaningful to see
non-perturbative QCD property in such a gravity background.

Most works to see such a property of gauge theories have been performed 
for heavy
fundamental quarks, while an idea to add light flavor quarks 
has been recently
proposed by Karch and Katz~\cite{KK} for D3-D7 brane system in 
AdS$_5\times S^5$.
Several authors have extended this idea to various 10d gravity backgrounds
which have been proposed for various gauge duals, and they have tried to
show meson spectrum,
chiral symmetry breaking and other related subjects
in several appropriate gravity backgrounds~\cite{KMMW,KMMW2,Bab,ES,SS,NPR}.

\vspace{.3cm}
Our purpose here is to study non-perturbative properties of QCD 
from gravity side through a model, which includes a non-trivial dilaton,
by introducing D7-brane probe according to the idea of Karch-Katz.
We solve the embedding problem of
the D7-brane in some backgrounds to study the chiral symmetry breaking
of the flavor quark. We consider firstly
a supersymmetric background given in~\cite{KS2,LT}. 
In this background, there is no singularity, and
the quark confinement has been assured for heavy fundamental
quark~\cite{KS2,LT}. 
Here we concentrate on the light flavor quarks to
study chiral symmetry breaking and 
inter-quark potential. Then we extend the same analysis to a non-supersymmetric
case.

Quark confinement is seen in both supersymmetric and nonsymmetric cases.
While, as for the chiral symmetry, we find that it is not broken 
in the supersymmetric case, and we need a non-supersymmetric background.
Such an example is shown and we 
assured the spontaneous chiral symmetry breaking (SCSB) for
the model given here. 

\vspace{.2cm}
In section 2, we give the setting of our model and the embedding of D7 brane
in both supersymmetric and non-symmetric backgrounds to study
the breaking of the chiral symmetry.
In section 3, the quark-antiquark
potential for flavor quarks is studied through the Wilson loop estimation 
for the two backgrounds, and
the summary is given in the final section.
\section{Background geometry and embedding of D7 brane}

The D7 brane embedding is studied for
two types of backgrounds, supersymmetric and non-symmetric one, 
to see the chiral symmetry
breaking for the light flavor quarks introduced through D7 brane probe.

\subsection{Supersymmetric background}

We consider the $ISO(1,3) \times SO(6)$ symmetric 
solution given in ~\cite{KS2,LT} for 10d IIB model.
This solution is supersymmetric and it has no singularity
in the bulk, so we can study the dual gauge theory through 
semi-classical approach of bulk gravity. In the present case, the dual
gauge theory for this background 
preserves ${\cal N}=2$ supersymmetry. The solution can be written 
in the string frame and taking $\alpha'=g_s=1$, as follows,
\beq 
ds^2_{10}=G_{MN}dX^{M}dX^{N}= e^{\Phi/2}
\left(
{r^2 \over R^2}\eta_{\mu\nu}dx^\mu dx^\nu +
\frac{R^2}{r^2} dr^2+R^2 d\Omega_5^2 \right) \ , 
\label{SUSY-sol}
\eeq 
\beq
e^\Phi= \left( 1+\frac{q}{r^4} \right) \ , \quad \chi=-e^{-\Phi}+\chi_0 \ ,
\label{dilaton}
\eeq
where $M,~N=0\sim 9$, $x^{\mu}=X^{\mu} (\mu,~\nu=0\sim 3)$, 
$R^4=4 \pi N g_s$ and $q$ is a constant. $\Phi$ and $\chi$
denote the dilaton and the axion respectively, and the self-dual five form
$F_{\mu_1\cdots \mu_5}$ is given as in \cite{KS2,LT}. 
And other field configurations are not used here.
This solution connects two asymptotic geometries, AdS$_5\times S^5$
and flat space-time, respectively, in UV ($r=\infty$) 
and IR ($r=0$) limit~\cite{KS2,LT}.

\vspace{.3cm}
We introduce flavor quark by embedding a D7 brane probe which
lies in $\left\{x^{\mu}, X^4\sim X^7\right\}$ directions. Here we
rewrite the 6d geometry as 
$\sum_{M=4}^{9} (dX^M)^2=dr^2+r^2d\Omega_5^2=d\rho^2+\rho^2d\Omega_3^2
+(dX^8)^2+(dX^9)^2$, where $\rho^2=\sum_{M=4}^7(X^M)^2$. Then
$r^2=\rho^2+(X^8)^2+(X^9)^2$.
The brane action for the D7-probe is 
\beq
S_{\rm D7}= -\tau_7 \int d^8\xi \left(e^{-\Phi} \sqrt{\cal G} 
      +{1\over 8!}\epsilon^{i_1\cdots i_8}A_{i_1\cdots i_8}\right) \ ,
\label{D7-action}
\eeq
where ${\cal G}=-{\rm det}({\cal G}_{i,j})$, $i,~j=0\sim 7$. 
${\cal G}_{ij}= \partial_{\xi^i} X^M\partial_{\xi^j} X^N G_{MN}$
and $\tau_7$ represent the induced metric and 
the tension of D7 brane respectively. Here we consider the case
of zero $U(1)$ gauge field on the brane, 
but we notice that the eight form potential $A_{i_1\cdots i_8}$, 
which is Hodge dual to the axion, couples to the 
D7 brane minimally. We obtain the eight form potential $A_{(8)}$
as $F_{(9)}=dA_{(8)}$ in terms of the Hodge dual field strength $F_{(9)}$ 
\cite{GGP}.
By taking the canonical gauge, $\xi^i=X^i$, we fix the embedding by
solving the equation of motion for the fields $X^8(\xi)$ and $X^9(\xi)$ 
under the ansatz, $X^9\equiv w(\rho)$ and $X^8=0$, without
loss of generality.
Then the induced metric and the action (\ref{D7-action}) are reduced as  
\beq
ds^2_8=e^{\Phi/2}
\left(
{r^2 \over R^2}dx^\a dx_\a + 
\frac{1+(w')^2}{r^2}R^2 d\rho^2 
+\frac{\rho^2}{r^2}R^2 d\Omega_3^2 \ .
 \right)
\eeq
\beq
S_{\rm D7-S} =-\tau_7~\int d^8\xi  \sqrt{\epsilon_3}\rho^3
\left( 1+\frac{q}{r^4} \right)\left(
   \sqrt{ 1 + (w')^2 }-1\right)
\ ,
\label{SUSY-w9}
\eeq
where $w'=dw/d\rho$ and $\epsilon_3$ is the determinant of three sphere.
And we obtain the following equation
\beq
 w''+ {3\over \rho}w'(1+(w')^2) 
    + {4q\over r^2(r^4+q)}\left\{(w-\rho w')(1+(w')^2)-w(1+(w')^2)^{3/2}
        \right\} =0, \label{w9-eq-super}
\eeq
The solution $w$ determines the embedding of the D7-brane, and the
problem of the chiral symmetry breaking is also 
solved from the viewpoint
of the dual gauge theory. The latter point is understood 
from the asymptotic form of the solution.
For $r \to \infty$ (i.e. $\rho \to \infty$), 
the solution $w(\rho)$ can be solved as 
\beq
   w(\rho) \sim m + {c \over \rho^2} \sim
    m + {c \over r^2}\ . 
\label{w9-b}
\eeq
Since the conformal dimension of $w$ is three~\cite{KK}, then
we can interpret $m$ and $c$ are quark mass and the vev of quark bilinear
$<\bar{\psi}\psi>$, respectively, from the gravity/gauge correspondence. 

In the same holographic context, we can give a comment on the 
form of the dilaton $e^\Phi$. 
It represents the running coupling of the dual gauge theory, 
and the
parameter $q$ can be interpreted as the gauge field condensate,
$\langle F_{\mu\nu}F^{\mu\nu} \rangle$. 
As for the relation between $q$ and $\langle F_{\mu\nu}F^{\mu\nu} \rangle$, 
further consideration from gauge theory side is given in \cite{LT}.
So we abbreviate them, but this parameter is the most important factor
of the present model.

\begin{figure}[htbp]
\begin{center}
\voffset=15cm
  \includegraphics[width=7cm,height=7cm]{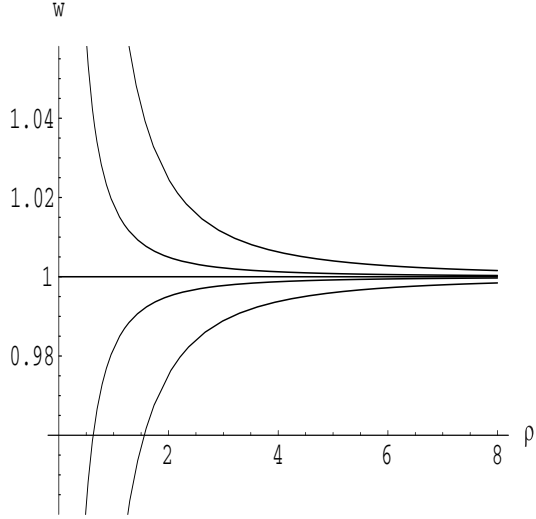} 
\caption{The solution $w$ as a function of $\rho$. 
Here we set $q=1$.  
The solutions are depicted for five cases of 
$c=-0.1,-0.02,~0,~0.02,~0.1$ with $m=1$ fixed. 
When $c$ is finite, the solutions are divergent at $\rho=0$. 
The regular solution is the mass only solution $w(\rho)=m$ represented by 
the horizontal line. 
\label{graph-w9-rho-1} }
\end{center}
\end{figure}

\vspace{.3cm}
Equation (\ref{w9-eq-super}) yields constant solutions, $w(\rho)=m$, 
and non-constant ones with $c\neq 0$ for each $m$. The latter solutions are 
obtained numerically and are shown in the Fig.\ref{graph-w9-rho-1}.
But they should be abandoned by the two reasons; 
(i) their energies are always 
higher than that of the constant solution with the same $m$, and (ii)
they can not be interpreted from the AdS/CFT
context due to the lack of one-to-one correspondence of $\rho$ and $r$
\cite{Bab}. On the other hand, all the constant solutions are 
equally possible since their
energies are degenerate to zero, $-S_{\rm D7-S}= 0$.
Then we can say that SCSB does
not occur in the supersymmetric background.

This result is
reasonable. For supersymmetric background, there is no force
between D3 and D7 branes for $w'=0$, then the value of $w(\rho)$ can not
be changed from $w(\infty)=m_q$ when $\rho\to 0$, and $w(\rho)=m_q$ is
preserved up to $\rho=0$. Therefore, 
in order to realize SCSB,
it would be necessary to consider
a non-supersymmetric background
solution as shown in the next subsection.

\vspace{.3cm}
Before considering the non-supersymmetric background,
we comment on the supersymmetry for the solutions of $w'\neq 0$.
The background (\ref{SUSY-sol}) has $1/2$ supersymmetry of 
IIB theory and its Killing spinor
is given as ~\cite{KS2}
\beq
  \epsilon=e^{\Phi/4}({r^2\over R^2})^{1/4}\epsilon_0,
\eeq
where $\epsilon_0$ is a constant spinor which satisfies 
$i\sigma_2\otimes \Gamma_{0123}\epsilon_0=\epsilon_0$. Here $\Gamma_{0123}$
denotes the antisymmetrized product of $\Gamma_{A}$ which are 10d flat 
space $\Gamma$ matrices.
After embedding D7-brane
probe with the above solution for $w(\rho)$, the supersymmetry is
in general broken except for constant $w$. This is easily seen as follows.
When some supersymmetry remains, the above Killing spinor must satisfy
the following condition \cite{BKOP} due to the $\kappa$ symmetry of D7 brane,
\beq
  \Gamma \epsilon=\epsilon, \qquad \Gamma=i\sigma_2\otimes \Gamma_{(0)}
   \label{susy-condition}
\eeq
\beq
  \Gamma_{(0)}={1\over 8!\sqrt{\cal G}}\epsilon^{i_1\cdots i_8}
  \partial_{i_1}X^{M_1}\cdots \partial_{i_8}X^{M_8}\Gamma'_{M_1\cdots M_8}
\eeq
where $\Gamma'_{M_1\cdots M_8}$ is the totally antisymmetrized product of
$\Gamma'_{M}=E_{M}^{A}\Gamma_{A}$. In our case, in the Cartesian coordinate,
we obtain
\beq
  \Gamma_{(0)}={1\over \prod_{i=4}^{7}\sqrt{1+(X^iw'/\rho)^2}}
  \left(\Gamma_{(1)}+{w'\over \rho}\sum_4^7 X^i\Gamma_{(i)}\right)
\eeq
where
$\Gamma_{(1)}=\Gamma_{01234567}$ and $\Gamma_{(i)} (7\geq i\geq 4)$ is the one in which
the number $i$ is replaced by $9$ in $\Gamma_{(1)}$. When $w'=0$, the
condition of the supersymmetry is written as
\beq
   i\sigma_2\otimes \Gamma_{(1)}\epsilon_0=\epsilon_0
\eeq
and we could find $1/2$ supersymmetry of the original one. However, for
non-zero $w'$ we find that supersymmetry is completely broken.
So it is natural to consider a non-supersymmetric background to see
SCSB.

\subsection{Non-supersymmetric background}

Here we consider a non-supersymmetric solution \cite{KS-G,NO,GTU}
which is given without
changing the five form field and eliminating the axion, $\chi=0$, as,
\beq 
ds^2_{10}= e^{\Phi/2}
\left(
{r^2 \over R^2}A^2(r)\eta_{\mu\nu}dx^\mu dx^\nu +
\frac{R^2}{r^2} dr^2+R^2 d\Omega_5^2 \right) \ , 
\label{non-SUSY-sol}
\eeq 
\beq
 A(r)=\left((1-({r_0\over r})^8 \right)^{1/4}, \quad
 e^\Phi= \left({(r/r_0)^4+1\over (r/r_0)^4-1} \right)^{\sqrt{3/2}} \ .
\label{dilaton-2}
\eeq
This configuration has a singularity at the horizon $r=r_0$, and
the present semi-classical analysis can not be applied near this point.
So we avoid this point in the followings.

As for the the dilaton, $e^\Phi$ can be expanded as 
\beq
   e^\Phi \sim 1 + {q_{\rm{NS}} \over r^4} \; ,  \quad
   q_{\rm{NS}}=\sqrt{6} r_0^4 \; .
\eeq
As shown in the previous case, in the context of AdS/CFT, it would be possible
to interpret the parameter $q_{\rm NS}$ as 
the gauge-field condensate  $\langle F_{\mu\nu}F^{\mu\nu} \rangle$.  

The D7 brane is embedded as above, and the following brane action is
obtained,
$$ S_{\rm D7-NS}= -\tau_7~\int d^8\xi 
{\cal L}_{\rm NS}~~~~~~~~~~~~~~~~~~~~~~~~~~~~~~$$
\beq
=-\tau_7~\int d^8\xi  \sqrt{\epsilon_3}\rho^3
\left({(r/r_0)^4+1\over (r/r_0)^4-1} \right)^{\sqrt{3/2}}
\left(1-({r_0\over r})^8\right)\sqrt{ 1 + (w')^2 } 
\ .
\label{non-SUSY-w9}
\eeq
We 
could expect to find SCSB by solving the equation of motion for $w$,
\beq
 \partial_w {\cal L}_{\rm NS}-\partial_{\rho}{\partial {\cal L}_{\rm NS} 
\over \partial {w'}}=0.
   \label{equation-NS}
\eeq
\begin{figure}[htbp]
\begin{center}
\voffset=15cm
  \includegraphics[width=7cm,height=7cm]{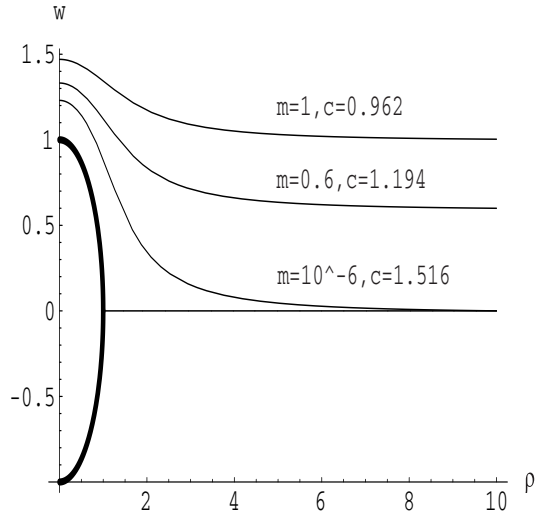} 
\caption{The solution $w$ as a function of $\rho$. 
Here, we take $r_0=1$, or $q_{\rm NS}=\sqrt{6}$.  
The non-trivial solutions are depicted for three cases of 
$m=10^{-6},~0.6,~1$. 
The curve for $m=10^{-6}$ agrees with the one in the chiral limit 
within the thickness of line. 
The thick-solid curve represents the horizon $w^2+\rho^2=r_0^2$. 
The horizontal line shows the trivial solution ($w(\rho)=0$) and the endpoint 
touches the horizon at $(\rho,w)=(1,0)$. 
\label{graph-w9-rho} }
\end{center}
\end{figure}
We consider the neighborhood of $\rho=0$ and set $w(0)'=0$
since the solution must be an even function of $\rho$. Then
Eq. (\ref{equation-NS}) is solved numerically 
for $w(\rho)$ by setting $w(0)'=0$ and $w(0) > r_0$; note that 
there exists a singularity at $r=\sqrt{\rho^2+w^2}=r_0$.  
Figure \ref{graph-w9-rho} shows the solutions for three values 
of $w(0)$, where $r_0=1$, or $q_{\rm NS}=\sqrt{6}$, is taken. 
Each solution yields a different set of $m$ and $c$.   
As for the solution of the $m=0$ limit, $c$ is finite. Thus, 
the spontaneous chiral symmetry breaking takes place in the present 
non-supersymmetric background. 
Other two solutions also show finite $<\bar{\psi}\psi>$. 
The solutions become flatter, as $m$ increases. 
Similar $\rho$ dependence of $w(\rho)$ is also seen in \cite{Bab} 
for a D7 insertion in the Constable-Myers background~\cite{CM} 
which is non-supersymmetric and has a non-constant dilation. 
The solutions which touch on the horizon are omitted here by the
reason mentioned above. A typical example of such solutions is 
the trivial solution $w(\rho)=0$, 
as shown in Fig. \ref{graph-w9-rho}. 
So the trivial solution, $w=0$, is abandoned
here differently from the supersymmetric case.

\vspace{.3cm}

\section{Quark-antiquark potential}

In this section, we study the static potential between a dynamical 
quark-antiquark pair in the above two backgrounds.

\subsection{Supersymmetric background}

In the case of $\mathcal{N}=2$ symmetric and constant $\Phi$, the potential
has been estimated in \cite{KMMW}.
And it has been shown that non-confinement Coulomb potential is seen
at large separation of the quarks. 
Here we show, for finite quark mass, 
the linear rising potential which has been shown for the
case of infinitely heavy quarks \cite{KS2,LT}. 

\vspace{.3cm}
Here we consider the supersymmetric case given above.
~The relevant part of the AdS$_5$ metric is written as
\beq
  ds^2=e^{\Phi/2}\left({r^2\over R^2}(-dt^2+dx^2)+{R^2\over r^2}dr^2\right).
\eeq
We consider a string whose endpoints lie on D7-brane at a distance $2L$
from one another. And the string is straddling the point $x=0$ about which
the profile is symmetric. Then the string action per unit time, namely
its energy, is given as
\beq
  E={1\over\pi}\int_0^{L}dx~n\sqrt{{r^4\over \lambda}+(r')^2},
\qquad n=\sqrt{1+{q\over r^4}}, \label{energy}
\eeq
where $\lambda=R^4$ denotes the 'tHooft coupling and $r'=dr/dx$; 
here we take $g_s=1$ for simplicity.
Since the "Lagrangian" is independent of x, we can introduce a constant
$u_0$ as
\beq
  \sqrt{\lambda}u_0^2= {n\over\sqrt{{r^4\over \lambda}+(r')^2}}r^4.
 \label{r0}
\eeq
From this, we obtain the following by introducing $y=r/u_0$,
\beq
  L={\sqrt{\lambda}\over u_0}\int_{y_{\rm min}}^{y_{\rm max}}{dy\over y^2}~
{1\over\sqrt{\tilde{q}+y^4-1}},
\qquad \tilde{q}={q\over u_0^4},
\eeq
where 
\beq
 y_{\rm max}=r_{\rm max}/u_0~, \quad 
 y_{\rm min}=(1-\tilde{q})^{1/4}~.
\eeq
Here we comment on the above upper 
bound $r_{\rm max}$. We should take it as $r_{\rm max}=w(0)>0$, 
however $w(0)=w(\infty)$ in the present case since $w(\rho)$ is a constant.
Then we can consider as $r_{\rm max}=2\pi m_q$ as in the AdS case
of $q=0$. 

\vspace{.2cm}
As for the energy, we obtain
\beq
  E={u_0\over \pi}\int_{y_{\rm min}}^{y_{\rm max}}{dy\over y^2}~
{\tilde{q}+y^4\over\sqrt{\tilde{q}+y^4-1}}.
\eeq

\begin{figure}[htbp]
\begin{center}
\voffset=15cm
  \includegraphics[width=9cm,height=7cm]{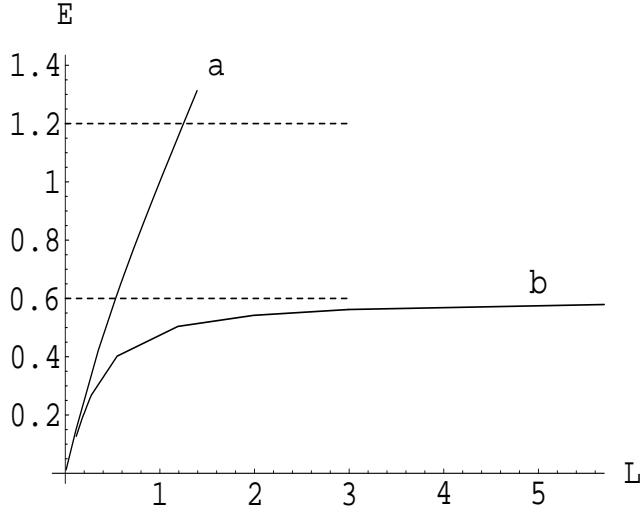} 
\caption{The energy $E$ of the string versus quark-antiquark
distances $2L$ are shown. 
The solid curve represents the one for (a) $q=5$, the confinement case, 
and 
for (b) $q=0$, non-confinement case. The normalization is free
and we set as $m_q=0.3,~ g_s=1$.
\label{mxigraph2}}
\end{center}
\end{figure}

\vspace{.3cm}
For $q=0$, the above formula are equivalent to the one given in \cite{KMMW},
and we find the Coulomb potential at long distance in this case. This means
that two quarks can be separated by very small energy
at infinitely long distance, $L\to\infty$,
where the energy approaches to $2m_q$ as shown in Fig.\ref{mxigraph2}.
Then quarks are not confined and could be free.

For
the case of $q>0$, the lower bound in the above integrals
is $(1-\tilde{q})^{1/4}$ and the dominant contribution is obtained from the
region of $\tilde{q}=q/u_0^4\sim 1$ or $u_0\sim q^{1/4}$. Near this region,
we obtain large $L$, and
we can estimate the above formulas as
\beq
  L=({\lambda^2\over q})^{1/4}{1\over 3y_{\rm min}^3}
\eeq
\beq
  E=2m_q+{\sqrt{q/\lambda}\over \pi}L
\eeq
This result expresses the linear potential and the effective string tension is
given as
\beq
  \tau={\sqrt{q/\lambda}\over \pi}
\label{string-tension}
\eeq
which is equivalent to the one given for the case of heavy quarks \cite{LT}.
So the energy of the bound state, in this confinement case,
exceeds $2m_q$ at some finite value of $L$, 
and increases with $L$ infinitely. 
However, in the present case, there are light dynamical quarks, then
the bound state would decay to two
mesons by a pair creation of quark and antiquark. The total energy of
newly created two mesons would be small when the distance between
quark and antiquark in those mesons is small enough. 
So the transition from large $L$ bound
state to light two mesons would be realized. 

\vspace{.3cm}
In the region of small $L$, which is realized for large $u_0$,
we could observe the similar potential
to the case of $q=0$ given in \cite{KMMW}.
This is easily understood by considering the fact
that the above formula
for $L$ and $E$ can be approximated by the one of $q=0$ for $q/u_0 << 1$.
We notice that we observe a linear potential again at very small $L$
with an effective string tension as given in \cite{KMMW}. The tension
is proportional to $m_q^2$ and is different from the one given above for
large $L$. This linear potential observed at small $L$
is caused by the strong coupling gauge
theory in the UV limit. This is then different from the 
case of real QCD which
is asymptotic free. The typical potential profile mentioned above
is shown in the Fig.\ref{mxigraph2} where two cases of $q=5$ and $q=0$ are shown.

\vspace{.3cm}
Finally, 
we comment on the case of electric condensate, 
${q}<0$. The integrant of the Equation (\ref{energy}) is rewritten as,
$\tilde{n}\sqrt{1/\lambda+(r')^2/r^4}$ and $\tilde{n}=\sqrt{q+r^4}$, and
$\tilde{n}$ has no finite minimum for ${q}<0$. Then the potential between
quark-antiquark does not show linear potential. It is easy to see this by
numerical estimation of $E$ as a function of $L$. As shown in the previous
section, instability appears in this case, and a peculiar behavior
will be seen also in the calculation of the Wilson loop. So we need a
care to study this quantity. We will discuss on this point in elsewhere.

\subsection{Non-supersymmetric case}

Next, we consider the non-supersymmetric solution, (\ref{non-SUSY-sol}) 
and (\ref{dilaton-2}). After the same
procedure as the supersymmetric case, we obtain the distance $2L_{\rm NS}$
and the energy $E_{\rm NS}$ for the light quark and anti-quark in terms of
an integral constant $H$, 
\beq
  H=e^{\Phi/2}({r\over R})^4{A(r)^3\over \sqrt{({r\over R})^4A(r)^2+(r')^2}},
\eeq
given as follows
\beq
  L_{\rm NS}=\int_{r_{\rm min}}^{r_{\rm max}}{dr\over r^2}~
{R^2\over A(r)\sqrt{F(r)-1}},
\qquad F(r)={e^{\Phi}\over H^2}A(r)^4({r\over R})^4, \label{l-NS}
\eeq
\beq
  E_{\rm NS}={1\over \pi}\int_{r_{\rm min}}^{r_{\rm max}}{dr\over R^2}~
{e^{\Phi}A(r)^3 r^2\over H\sqrt{F(r)-1}}.  \label{e-NS}
\eeq
Here $\Phi$ and $A(r)$ are given in (\ref{non-SUSY-sol}) 
and (\ref{dilaton-2}), and
the lower bound $r_{\rm min}$ of the integration is given by 
\beq
   F(r_{\rm min})=1.  \label{rmin}
\eeq
This is the middle point of the string connecting 
quark and anti-quark, and $r'=dr/dx$ should be zero at this point due to
the smoothness of the string configuration. Actually
we can see the equivalence of the equations 
$F(r_{\rm min})=1$ and $r'(r_{\rm min})=0$.

\begin{figure}[htbp]
\begin{center}
\voffset=15cm
  \includegraphics[width=9cm,height=7cm]{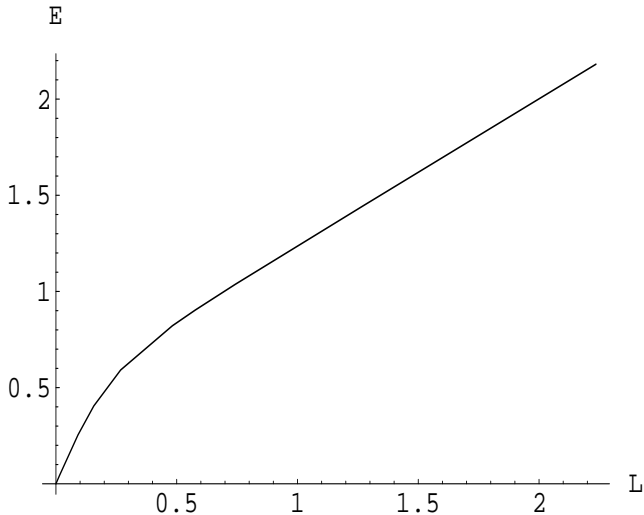}
\caption{The energy of the string versus quark-antiquark
distance is shown for non-supersymmetric case, (\ref{e-NS}) 
and (\ref{l-NS}). 
Here we set $m_q=3/(2\pi),~ g_s=1$, $r_0=1$ and $R=1$. 
\label{EL-NS}}
\end{center}
\end{figure}

\vspace{.3cm}
From the above formula, we can observe the linear potential for large $L$
as pointed out by \cite{KS-G}. 
In Fig.\ref{EL-NS}, an example of the numerical estimation for the above
$E$ and $L$ is shown.

\section{Summary}

A supersymmetric and a non-supersymmetric
background deformed by the dilaton are studied in the context of 
gauge/gravity correspondence by embedding a D7 brane. 
The supersymmetric background considered here
is dual to the $\mathcal{N}=2$ supersymmetric Yang-Mills theory
with gauge field condensate. In this case, we find
the energy minimum 
configuration of the embedded D7 brane
which is a flat plane in the eight
dimensional space-time. This solution implies zero vev of flavor quark
bilinear $\langle\bar{\psi}\psi\rangle$, then the chiral symmetry is preserved.
We find other non-flat solutions with $\langle\bar{\psi}\psi\rangle\neq 0$. 
However, these solutions break supersymmetry and have higher energies 
than the one of the supersymmetric flat solution for any quark-mass case.
Then we can say that the chiral symmetry is preserved in this background.

And, for any finite quark mass, we find the quark confinement 
in this case by estimating 
the Wilson-loop. Namely, we could find a linear potential for 
large separation of quark and anti-quark.
For the case of electric gauge-field condensate, the background has
a singular point where the action of the embedded D7 changes its sign. 
This implies that the available region for AdS/CFT analysis is bounded
by the magnitude of the electric field condensate, and we couldn't find 
the quark confinement in this state.

\vspace{.3cm}
As a result, in order to find a chiral symmetry breaking,  
we must consider non-supersymmetric background or an D7 embedding
which breaks supersymmetry. Here we examined a non-supersymmetric
background of IIB model, and
we could find a chiral symmetry breaking.  
And the inter-quark potential given by the Wilson loop shows the
confining force. However,
in our non-supersymmetric background configuration, there is a singular point.
So we must 
make the analysis out of this singularity.
One should notice that both results, CSB and the quark confinement, are obtained 
out of the singular region.

\vspace{.3cm}
\section*{Acknowledgments}

The authors are very grateful to M. Tachibana 
for useful discussions and comments throughout this work 
and also to N. Maru for useful 
discussions at the early stage of this work. 
K. G thanks C. Nunez for
useful and inspired discussions.
This work has been supported in part by the Grants-in-Aid for
Scientific Research (13135223, 14540271)
of the Ministry of Education, Science, Sports, and Culture of Japan.


\newpage
\end{document}